\documentclass[cits]{PoS}

\def\BS{$B_s^0$}

\def\BD{$B^0$}

\def\BRBSMUMU{${\cal B}(B_s^0 \rightarrow \mu^+ \mu^-)$}

\def\BUJPSIK{$B^+ \rightarrow J/\psi K^+$}

\def\BHH{$B \rightarrow h^+ h'^-$}
\def\BSMUMU{$B_s^0 \rightarrow \mu^+ \mu^-$}

\def\FULLOBSLIMITBR{${\cal B}(B_s^0 \rightarrow \mu^+ \mu^-) < 5.1 \times 10^{-8}  (4.2 \times 10^{-8}) $}

\def\BNN{$\beta$}

\title{Rare $B$ decays at the Tevatron}

\ShortTitle{Rare $B$ decays at the Tevatron}

\author{\speaker{Masato Aoki}\thanks{On behalf of the CDF and D0 Collaborations.}\\
  Fermi National Accelerator Laboratory\\
  E-mail: \email{masato@fnal.gov}\\
}

\abstract{
  Studying flavor changing neutral current transitions provides important information that helps searches for physics beyond the standard model.
  In this paper we report on recent measurements of these transitions using data collected by the CDF and D0 experiments at the Tevatron $p\bar{p}$ collider,
  including world-leading limits on the branching fraction of the decay $B^0_{(s)} \to \mu^+ \mu^-$,
  a forward-backward asymmetry measurement in $B^0 \to K^{*0} \mu^+ \mu^-$ and $B^+ \to K^+ \mu^+ \mu^-$ decays which is consistent and competitive with best $B$-factories results,
  and the first observation of the decay $B^0_s \to \phi \mu^+ \mu^-$.
}

\FullConference{Flavor Physics and CP Violation - FPCP 2010\\
		May 25-29, 2010\\
		Turin, Italy}

\begin{document}

\section{Introduction}
A flavor changing neutral current (FCNC) process is an apparent transition between quarks of different flavor but equal charge.
In the standard model (SM), these processes occur through higher order loop diagrams.
The SM branching fraction for such FCNC decays is quite small.
It is interesting to study such rare decays because the contribution from physics beyond the SM may be sizable.
In order to be able to observe rare heavy flavor decays it is essential to produce a sufficient number of $B$ hadrons.
At the Tevatron $p\bar{p}$ collider, various types of $B$ hadrons can be produced with large production cross sections of ${\cal O}(100~\mu{\rm b})$~\cite{Acosta:2004yw}.
However, the inelastic $p\bar{p}$ cross section is $10^{3}$ times larger than that of $B$ hadron production.
Interesting events have to be extracted from a high track multiplicity environment.
Therefore, detectors need to have very good tracking, vertex resolution, wide acceptance, good particle identification and highly selective triggers.

In this paper, we report on recent results from studies on rare $B$ decays via FCNC processes,
$B^0_{(s)} \to \mu^+ \mu^-$, $B^+ \to K^+ \mu^+ \mu^-$, $B^0 \to K^{*0} \mu^+ \mu^-$ and $B^0_s \to \phi \mu^+ \mu^-$,
performed by the CDF and D0 collaborations.

\section{$B^0_{(s)} \to \mu^+ \mu^-$}
The FCNC $B$ decays with dimuon in the final state are highly
suppressed by the helicity factor, $(m_\mu/m_B)^2$.
In the SM, the branching fraction of $B^0_s \to \mu^+ \mu^-$ decays is
$(3.6 \pm 0.3)\times 10^{-9}$~\cite{Buras:2009us}.
$B^0 \to \mu^+ \mu^-$ decay is further suppressed by the ratio of CKM matrix
elements, $|V_{td}/V_{ts}|^2$~\cite{Cabibbo,Kobayashi:1973fv},
giving a branching fraction in the SM of $(1.1 \pm 0.1) \times 10^{-10}$~\cite{Buras:2009us}.
The decay amplitude can be enhanced by several orders of magnitude in some extensions of the SM.
Since the predicted rate for these processes in the SM is beyond the
current experimental sensitivity at the Tevatron, the observation of
these decays would necessarily imply physics beyond the SM.  Improved
limits on the branching fraction can be used to set limits on the
parameter space of supersymmetric models and other new theories.
It is also important to measure the ratio of the branching fractions
between $B^0_s \to \mu^+ \mu^-$ and $B^0 \to \mu^+ \mu^-$ to test the Minimal Flavor Violation~\cite{Buras:2009us}.

\begin{figure}
  \begin{center}
    \includegraphics[scale=0.40]{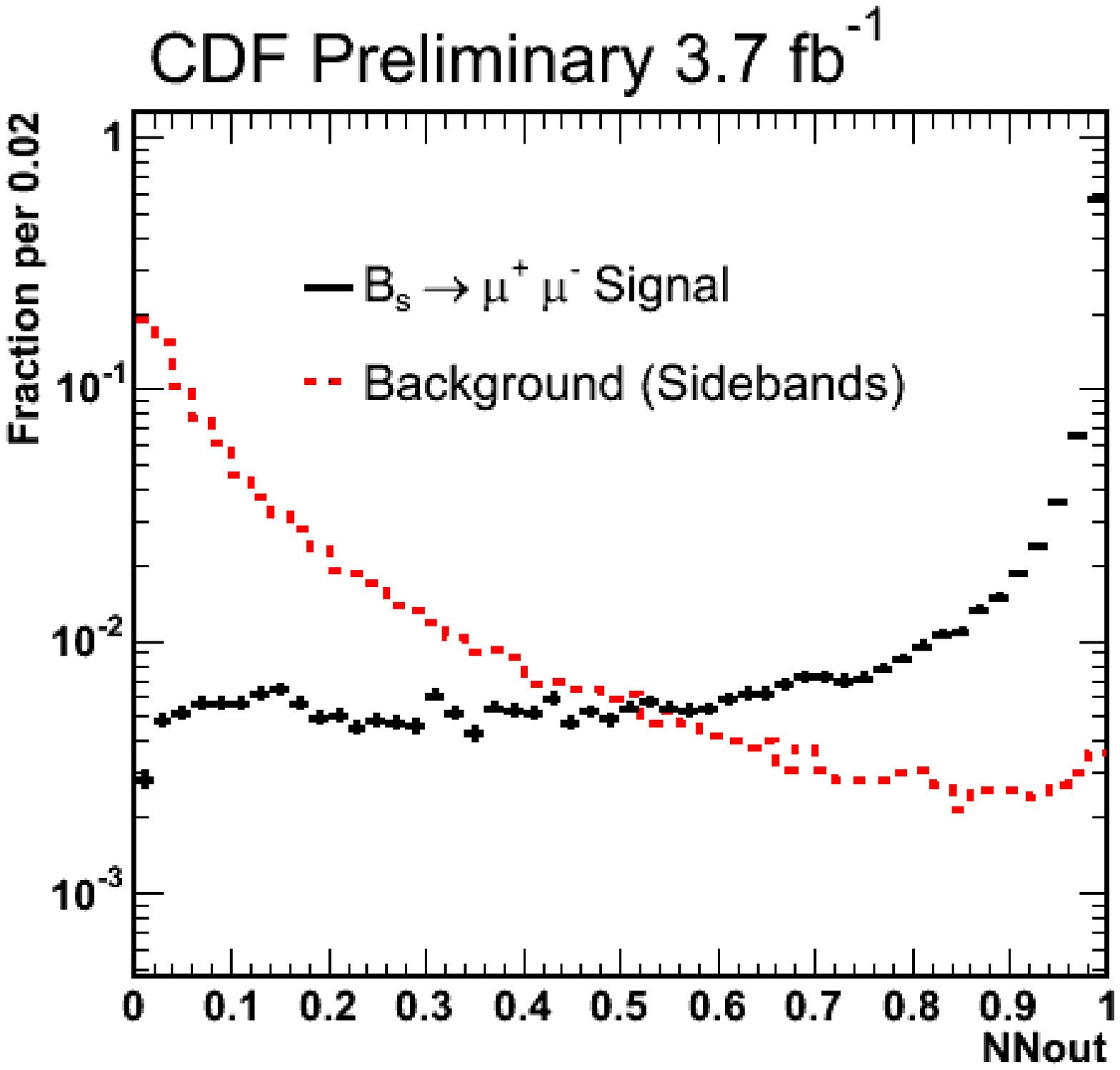}
    \includegraphics[scale=0.35]{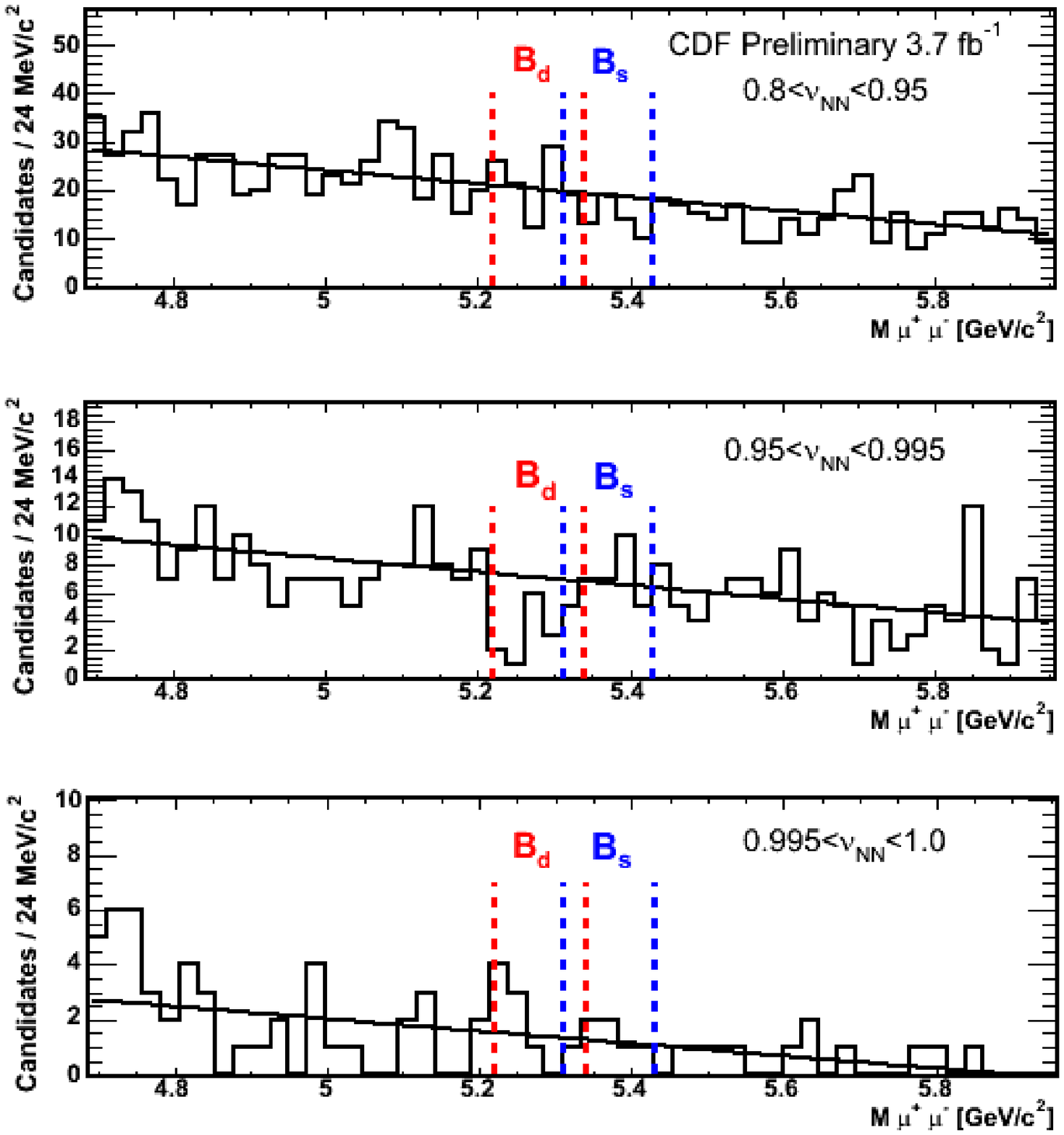}
    \caption{\label{fig:cdf_nn_mass} CDF data.
      Left: Distributions of the NN output for simulated $B^0_s \to \mu^+ \mu^-$ signal and sideband events in data.
      Right: The dimuon invariant mass distribution for events satisfying all selection criteria for the final three NN output ranges.
    }
  \end{center}
\end{figure}

CDF has performed a search for $B^0_{(s)} \to \mu^+ \mu^-$ with $3.7~{\rm fb}^{-1}$ of data~\cite{cdf_bsmumu}.
Their acceptance is increased compared to the previous analysis~\cite{cdf_bsmumu_pub}, in addition to increased luminosity,
by using events where muon candidates cross the mid-plane of the central outer tracker (COT)
where plastic inserts to maintain wire spacing create a zone with lower trigger efficiency.
This additional trigger acceptance increases the total trigger acceptance by 12\%.
Backgrounds from hadrons misidentified as muons are now suppressed by selecting muon candidates using a likelihood function.
This function tests the consistency of electromagnetic and hadronic energy with that expected for minimum ionizing particles
and the differences between extrapolated track trajectories and muon system hits.
In addition, backgrounds from kaons that penetrate through the calorimeter to the muon system, or decay in flight outside
the COT, are further suppressed by a loose selection based on the measurement of the ionization per unit path length, $dE/dx$.
After applying their baseline selections, a Neural Network (NN)  multivariate classifier is constructed using the following variables
to achieve further separation of signal from background:
(1) the measured proper decay time,
(2) the proper decay time divided by the estimated uncertainty,
(3) the 3D opening angle between the dimuon flight direction vector and the displacement vector between the interaction point and the dimuon vertex,
(4) the $B$-candidate track isolation,
(5) the $B$-candidate transverse momentum,
(6) and the transverse momentum of the lower momentum muon candidate.
The NN is trained using background events sampled from the sideband regions and signal events generated with a CDF simulation.
The NN output distributions of $B^0_s$ signal and sideband background events are shown in Fig.~\ref{fig:cdf_nn_mass}.
The search is performed in a two dimensional grid in dimuon mass $m_{\mu\mu}$ and NN space to improve the sensitivity.
The expected background is obtained by summing contributions from the combinatorial continuum and from $B \to h^+ h'^-$ decays ($h^+$ and $h'^-$ represent a charged kaon or pion),
which peak in the signal invariant mass region and do not occur in the sidebands.
The contribution from other heavy-flavor decays is negligible.
The combinatorial background is estimated by linearly extrapolating from the sideband region to the signal region.
The $B \to h^+ h'^-$ contributions are about a factor of ten smaller than the combinatorial background.
The dimuon invariant mass distributions for three different NN output ranges are shown in Fig.~\ref{fig:cdf_nn_mass}.
The limit on the branching fraction is computed by normalizing to the number of reconstructed \BUJPSIK\ events.
A sample of about $19,700$ \BUJPSIK\ events is collected to serve as a normalization channel using the same baseline requirements.
CDF extracts new upper limits ${\cal B}(B^0_s \to \mu^+ \mu^-) < 4.3 \times 10^{-8} (3.6 \times 10^{-8})$ and
${\cal B}(B^0 \to \mu^+ \mu^-) < 7.6 \times 10^{-9} (6.0 \times 10^{-9})$ at the 95\%~(90\%)~C.L., which are currently the world's best upper limits for both processes.

\begin{figure}
  \begin{center}
    \includegraphics[scale=0.35]{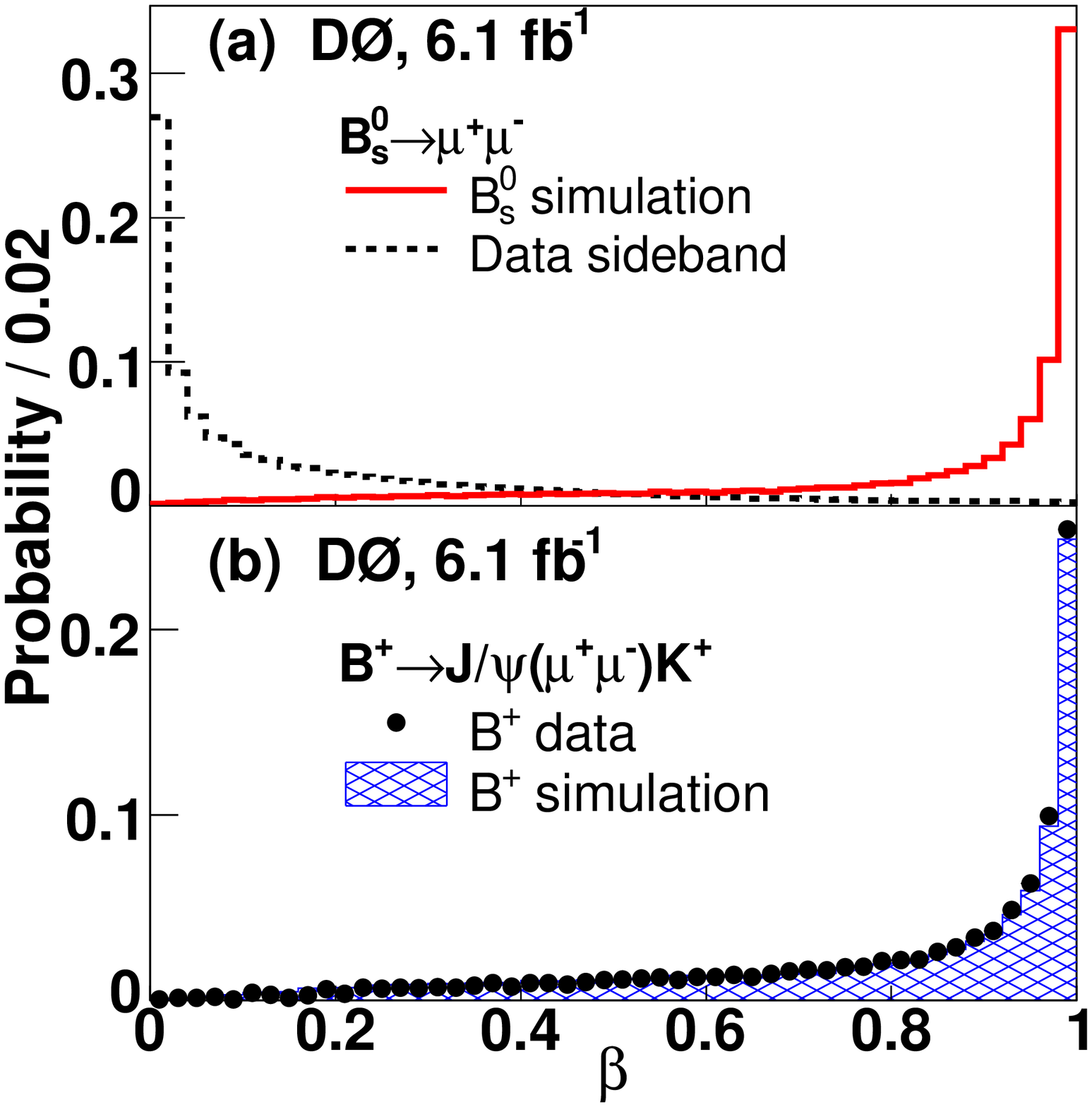}
    \includegraphics[scale=0.35]{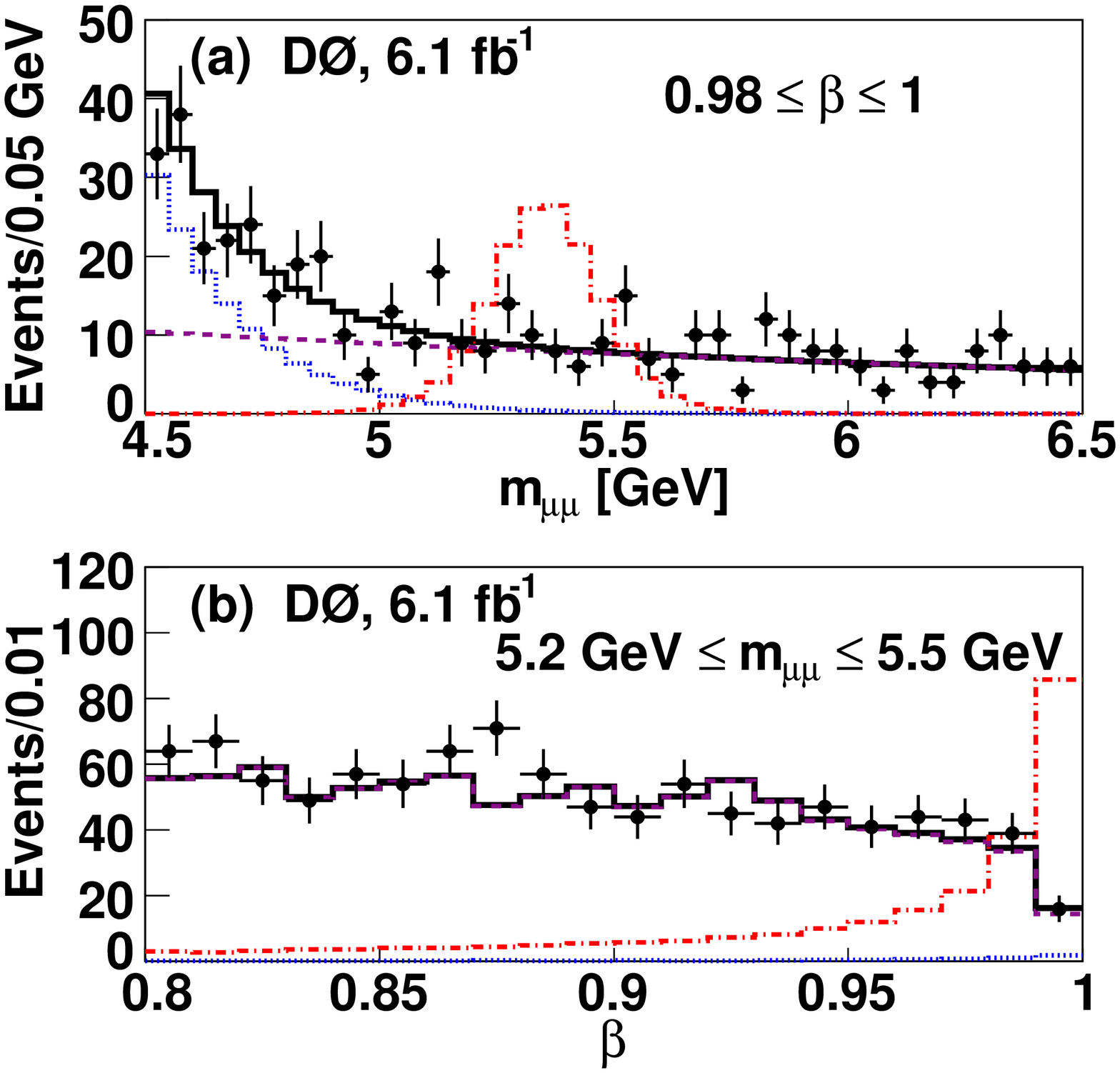}
    \caption{\label{fig:d0_nn_mass} D0 data.
      Left: Distributions of \BNN: (a) \BSMUMU\ signal and sideband events, (b) \BUJPSIK\ data and simulation.
      Right: The distribution of $m_{\mu\mu}$ in the highest sensitivity $\beta$ region (a),
      and the distribution of $\beta$ in the highest sensitivity $m_{\mu\mu}$ region (b)
      for data (dots with uncertainties), expected background distribution (solid line),
      and the SM signal distribution multiplied by a factor of 100 (dotted-dashed line).
      The dimuon background contributions from the $B(D) \to \mu^+ \nu X, {\bar B}(\bar{D}) \to \mu^- \bar{\nu} X'$ decays (dashed line)
      and the $B \to \mu^+ \nu \bar{D}, \bar{D} \to \mu^- \bar{\nu} X$ decays (dotted line) are also shown.
    }
  \end{center}
\end{figure}

D0 has also performed a search for \BSMUMU\ using $6.1~{\rm fb}^{-1}$ of data~\cite{d0_bsmumu}.
The event selection criteria are similar to those of CDF.
The muon selection has been updated with respect to the previous analysis~\cite{Abazov:2007iy},
yielding 10\% higher acceptance while keeping the fraction of misidentified muons below 0.5\%.
To further suppress the background, the following discriminating variables are used:
(a) the transverse momentum of the \BS\ candidate,
(b) the 3D opening angle,
(c) the transverse decay length significance,
(d) the decay vertex fit $\chi^2$,
(e) the smaller impact parameter significance of the two muons,
(f) and the smaller transverse momentum of the two muons.
A Bayesian Neural Network (BNN)~\cite{bnn1,bnn3} multivariate classifier with the above variables is constructed to distinguish signal events from background.
The BNN is trained using background events sampled from the sideband regions
and simulated signal events.
The distributions of the BNN output \BNN\ for the \BS\ signal and the sideband events as well as the \BUJPSIK\ control sample are shown in Fig.~\ref{fig:d0_nn_mass}.
The \BSMUMU\ signal region is defined to be $0.9 \leq {\rm \beta} \leq 1.0$ and $5.0~{\rm GeV} \leq m_{\mu\mu} \leq 5.8~{\rm GeV}$.
Two-dimensional (2D) histograms of $m_{\mu\mu}$~vs.~$\beta$, dividing the signal region into several bins, are prepared
to improve the sensitivity relative to using a single bin.
The dominant source of background dimuon events is from decays of heavy flavor hadrons in $b\bar{b}$ or $c\bar{c}$ production.
To study this background contribution, inclusive dimuon Monte Carlo samples with {\sc pythia}~\cite{pythia} generic QCD processes are generated.
The dimuon background events can be categorized by two types:
(i) $B(D) \to \mu^+ \nu X, {\bar B}(\bar{D}) \to \mu^- \bar{\nu} X'$ double semileptonic decays where the two muons originate from different $b(c)$ quarks,
yielding dimuon masses distributed over the entire signal region, 
and (ii) $B \to \mu^+ \nu \bar{D}, \bar{D} \to \mu^- \bar{\nu} X$ sequential semileptonic decays,
resulting in $m_{\mu\mu}$ predominantly below the $B$ hadron mass.
The simulated dimuon mass distributions for both background sources are parametrized using an exponential function
to estimate the number of background events in the signal region
after fitting the dimuon mass in the data sideband regions in each \BNN\ bin.
The uncertainty on this background estimate is dominated by the statistical uncertainty of the sideband sample ($10\%\sim35\%$).
The \BHH\ background contribution is negligible.
The limit on the branching fraction is computed by normalizing to the number of reconstructed \BUJPSIK\ events.
A sample of about $46,800$ \BUJPSIK\ events is selected using the same baseline selections.
The observed distributions of dimuon events in the highest sensitivity region are shown in Fig.~\ref{fig:d0_nn_mass}.
The observed number of events is consistent with the background expectations.
The resulting limit is \FULLOBSLIMITBR\ at the 95\%~(90\%)~C.L.
The simulated mass resolution of the D0 detector for the \BSMUMU\ is $\approx120~{\rm MeV}$
and is therefore insufficient to readily separate \BS\ from \BD\ leptonic decays.
In this analysis, it is assumed that there are no contributions from $B^0 \rightarrow \mu^+ \mu^-$ decays,
since this decay is suppressed by $|V_{td}/V_{ts}|^2 \approx 0.04$.

Aside from the background uncertainty, the largest uncertainty of 15\% common to CDF and D0 \BSMUMU\ analyses
comes from the fragmentation ratio between $B^+$ and $B^0_s$~\cite{pdg2006}.
Both CDF and D0 observe no evidence of physics beyond the SM in $B^0_{(s)} \to \mu^+ \mu^-$ decays,
but they significantly improve the existing limits providing tighter constraints for the parameter space of several SM extensions.



\section{$B^0 \to K^{*0} \mu^+ \mu^-$, $B^+ \to K^+ \mu^+ \mu^-$ and $B^0_s \to \phi \mu^+ \mu^-$}
Another sensitive probe of non-SM physics accessible at the Tevatron is to measure the FCNC $b \to s \ell \ell$ semileptonic decays.
Although there is no helicity factor in these decays, the SM branching fraction is still small, of order ${\cal O}(10^{-6})$,
and there could a sizable contribution to the decay rate from physics beyond the SM.
In addition, the kinematic distributions could be modified.
The lepton forward-backward asymmetry and the differential branching fraction
as a function of dilepton invariant mass  in the decays $B \to K^* \ell \ell$
differ from the SM expectations in various extended models.
The amplitudes of the effective Wilson coefficients $C_7$, $C_9$, and $C_{10}$ may interfere with the contributions from non-SM particles,
where $C_7$, $C_9$, and $C_{10}$ represent the amplitudes from
the electromagnetic penguin, the vector electroweak, and the axial-vector electroweak contributions, respectively.
Among many $b \to s \mu^+ \mu^-$ decays, the exclusive channels $B^+ \to K^+ \mu^+ \mu^-$ and $B^0 \to K^{*0} \mu^+ \mu^-$ have been observed and studied at Belle~\cite{:2009zv} and BaBar~\cite{:2008ju}.
At the Tevatron it is also possible to search for analogous decays $B^0_s \to \phi \mu^+ \mu^-$ and $\Lambda^0_b \to \Lambda \mu^+ \mu^-$.

\begin{figure}
  \begin{center}
    \includegraphics[scale=0.25]{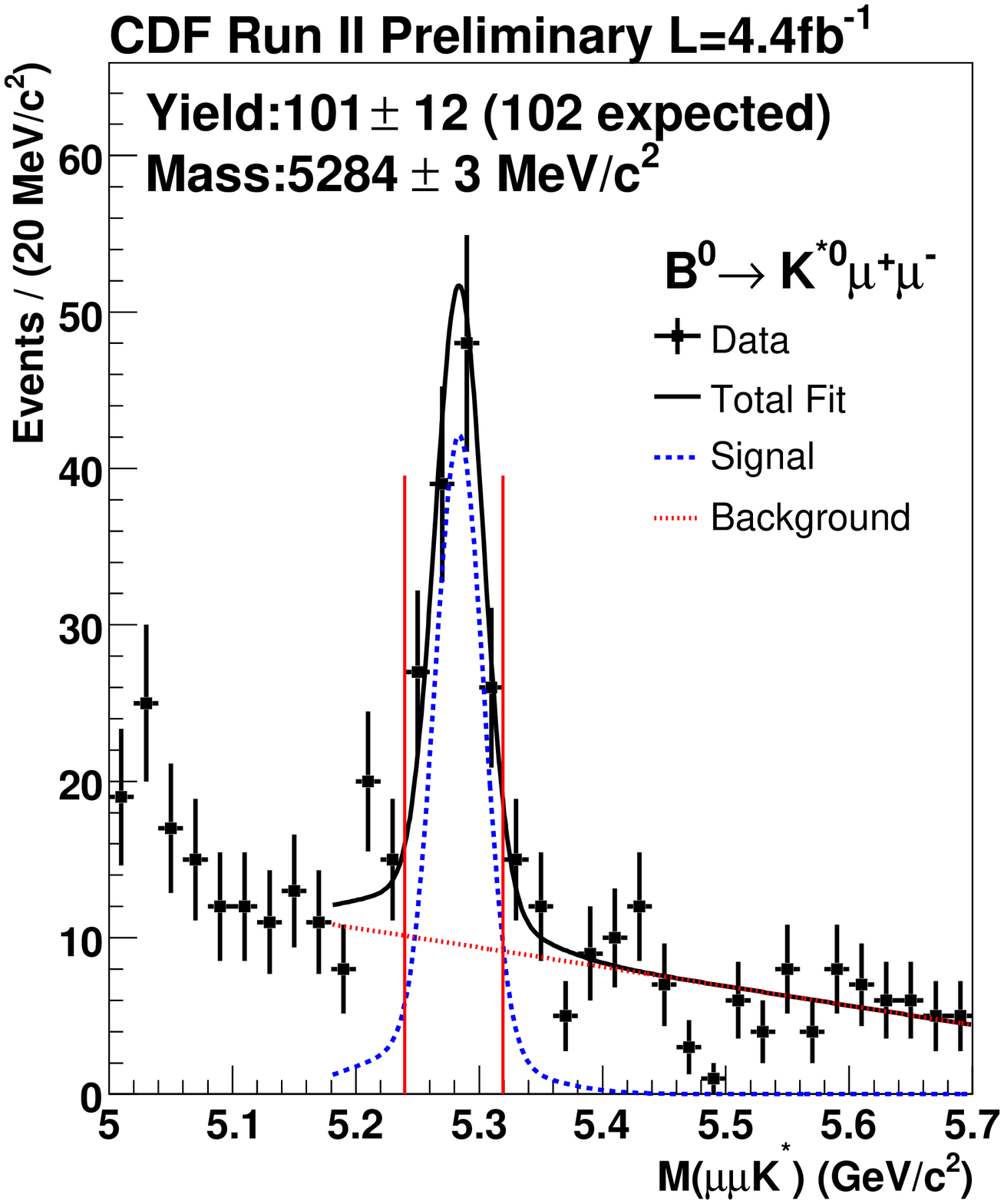}
    \hspace{1mm}
    \includegraphics[scale=0.25]{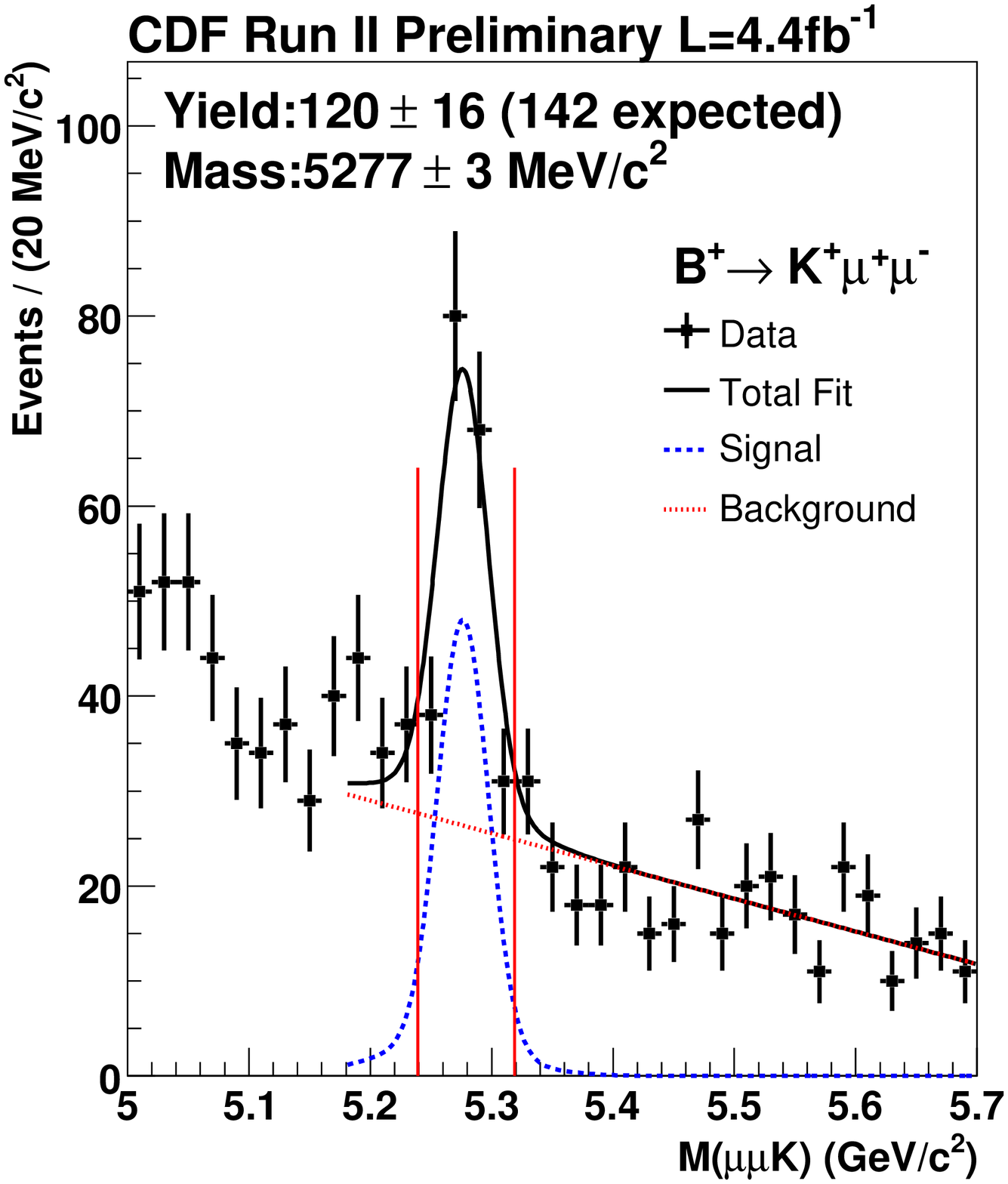}
    \hspace{1mm}
    \includegraphics[scale=0.25]{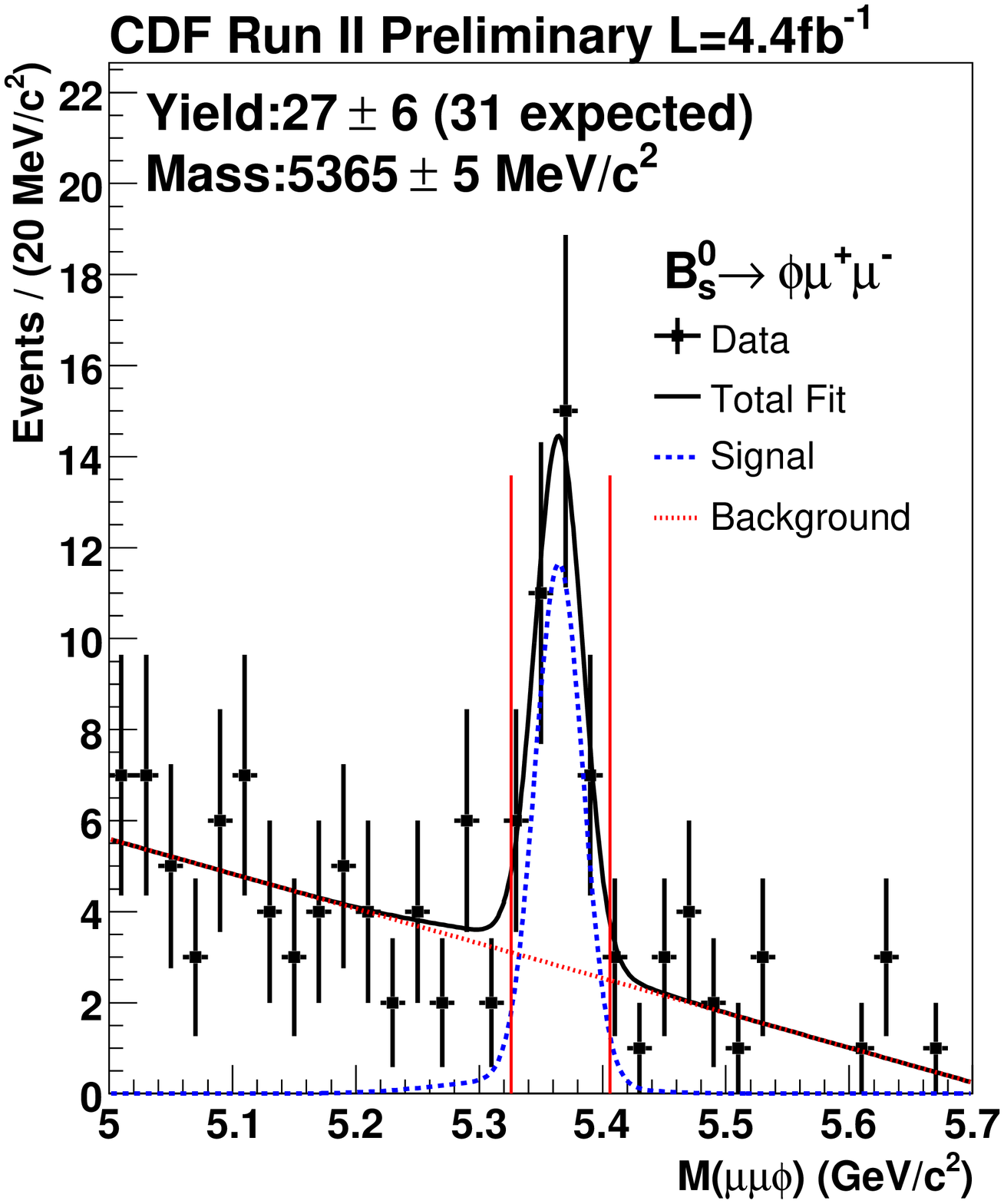}
    \caption{\label{fig:bmass_smm} CDF data. The $B$ invariant mass of $B^0 \to K^{*0} \mu^+ \mu^-$ (left), $B^+ \to K^+ \mu^+ \mu^-$ (middle) and $B^0_s \to \phi \mu^+ \mu^-$ (right).}
  \end{center}
\end{figure}

\begin{figure}
  \begin{center}
    \includegraphics[scale=0.3]{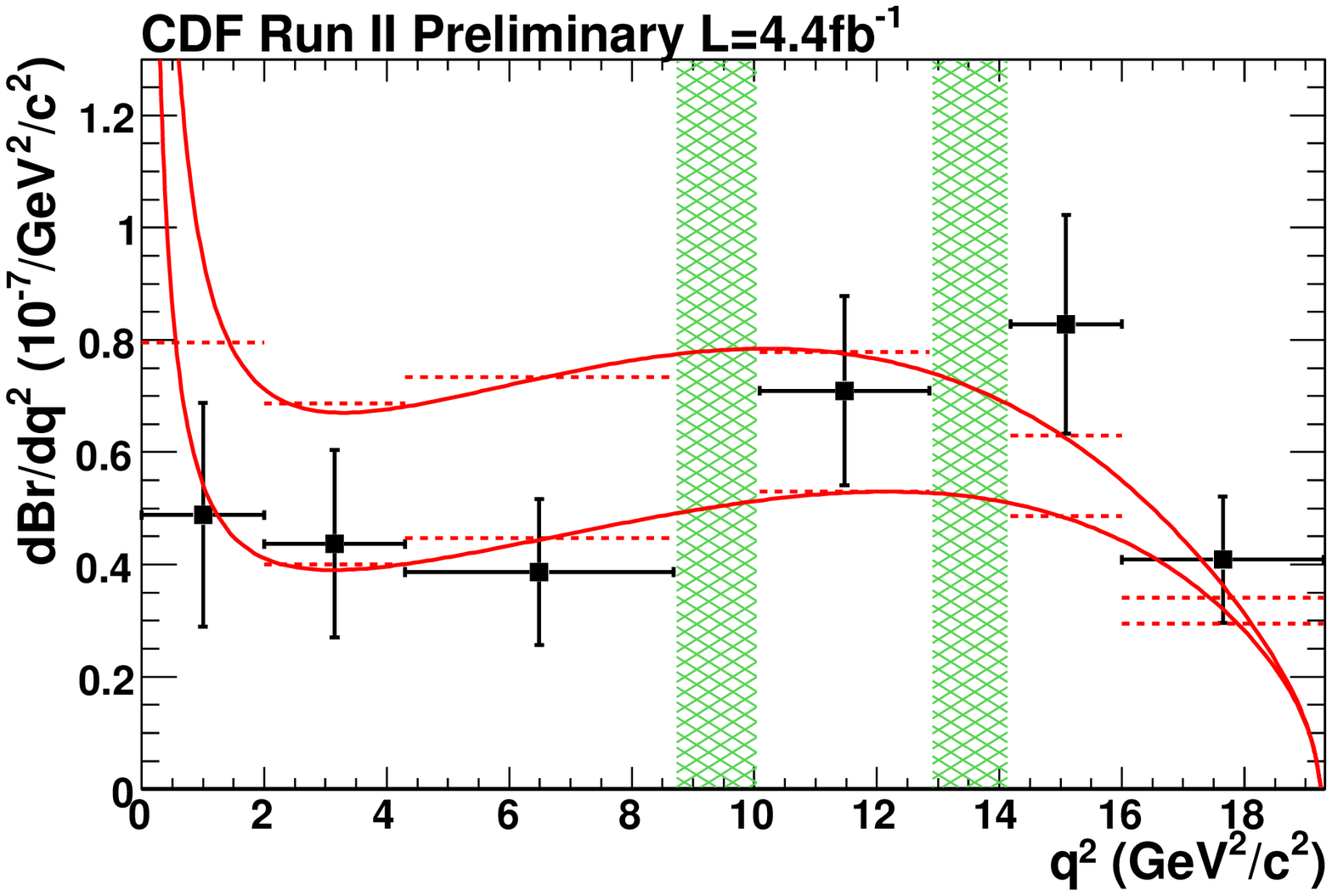}
    \hspace{1mm}
    \includegraphics[scale=0.3]{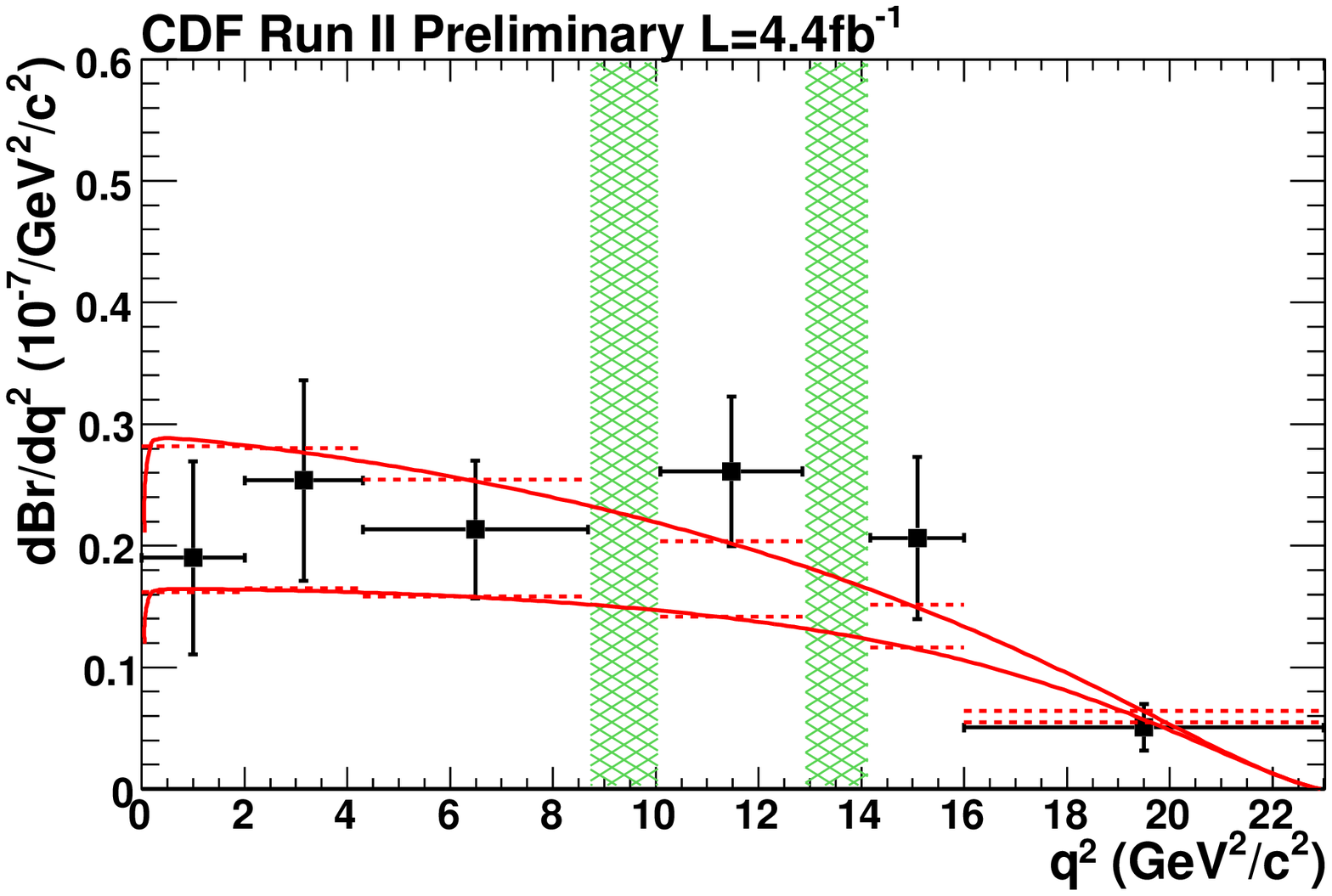}
    \caption{\label{fig:dbr} CDF data. Differential branching fraction of $B^0 \to K^{*0} \mu^+ \mu^-$ (left) and $B^+ \to K^+ \mu^+ \mu^-$ (right).
      The hatched regions are vetoed to prevent contributions from charmonium.
      The solid lines are the SM expectation, which use maximum- and minimum- allowed form factor.
      The dashed line is the averaged theoretical curve in each $q^2$ bin.}
  \end{center}
\end{figure}

\begin{figure}
  \begin{center}
    \includegraphics[scale=0.22]{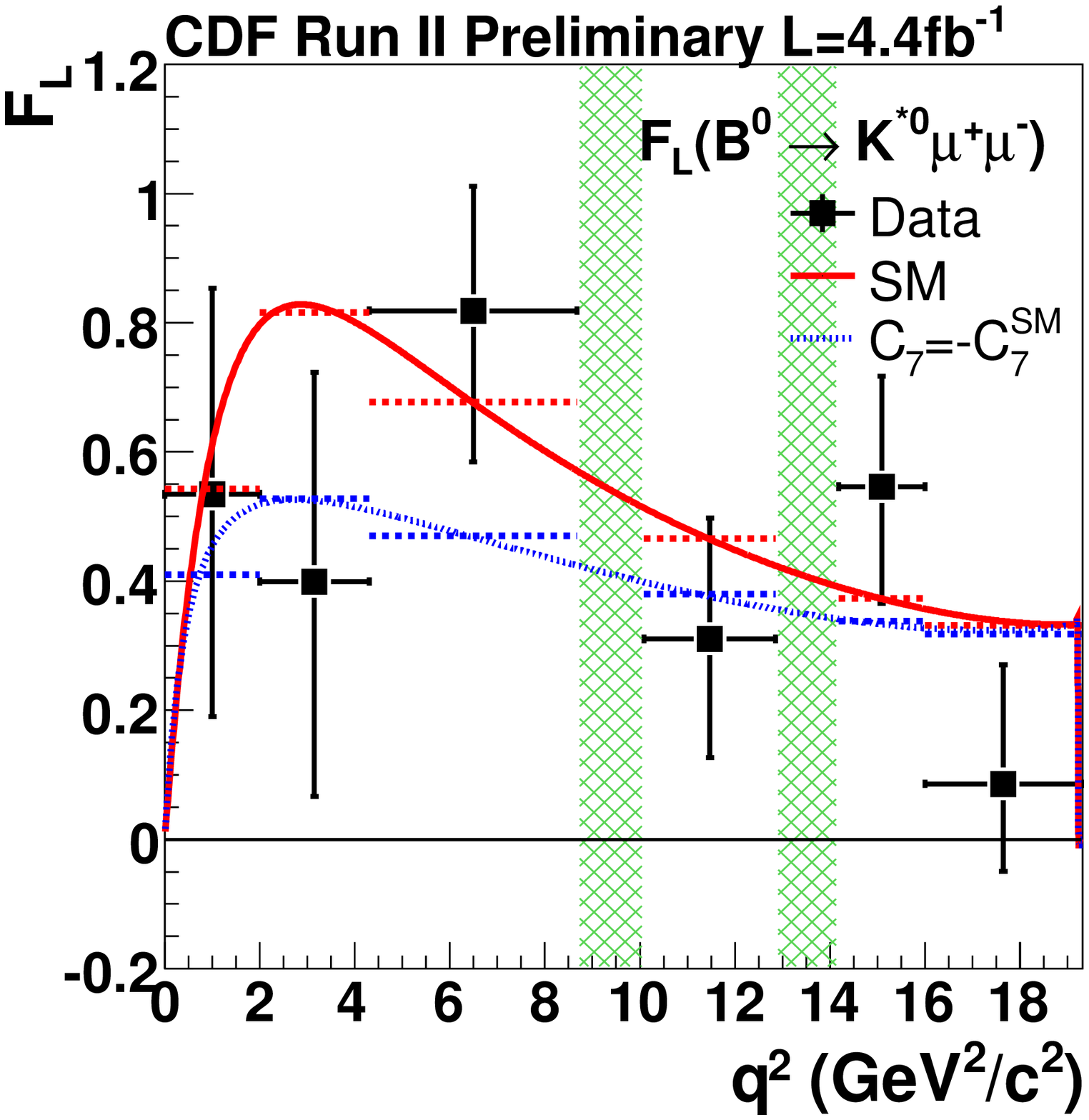}
    \includegraphics[scale=0.22]{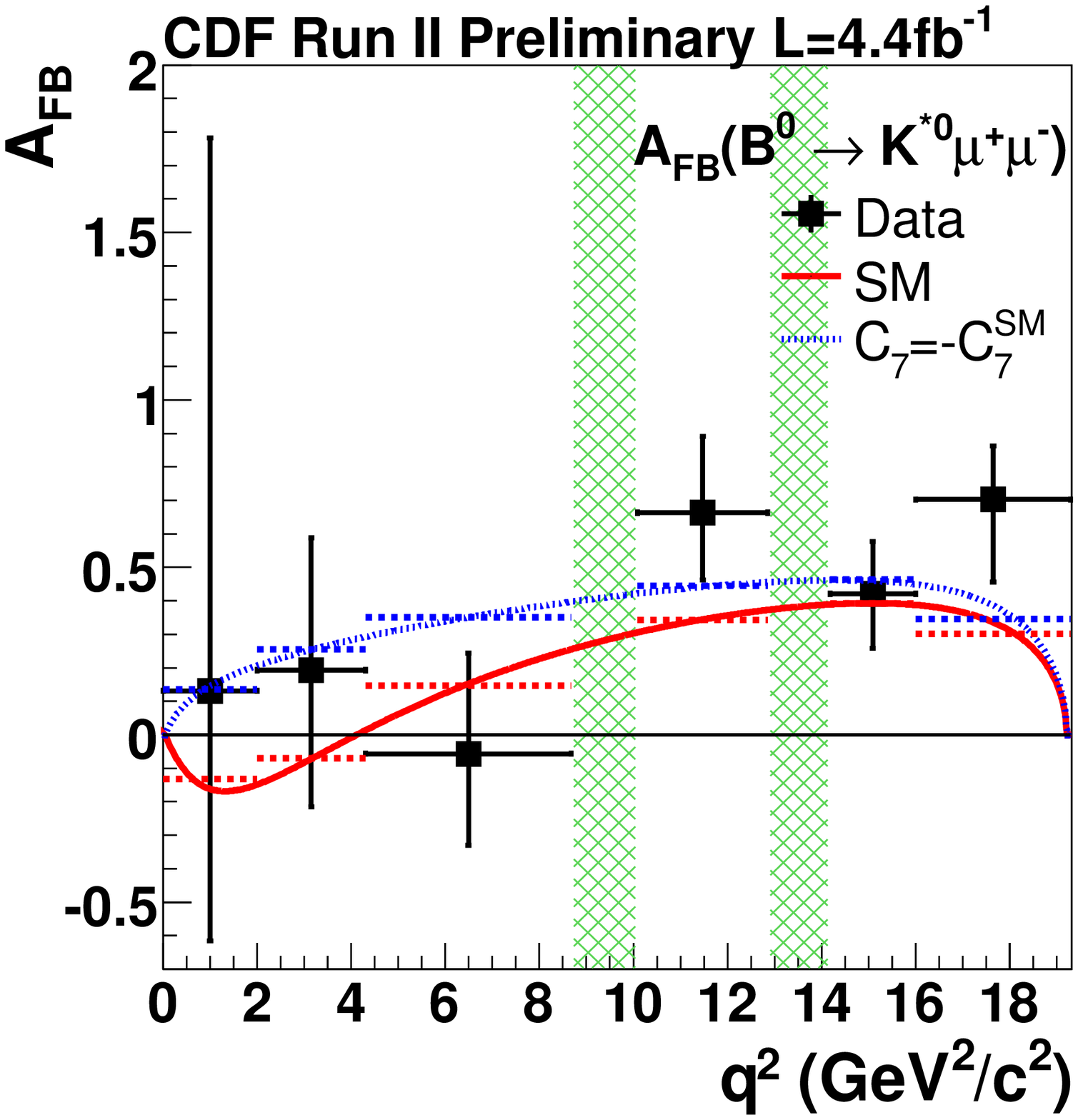}
    \includegraphics[scale=0.22]{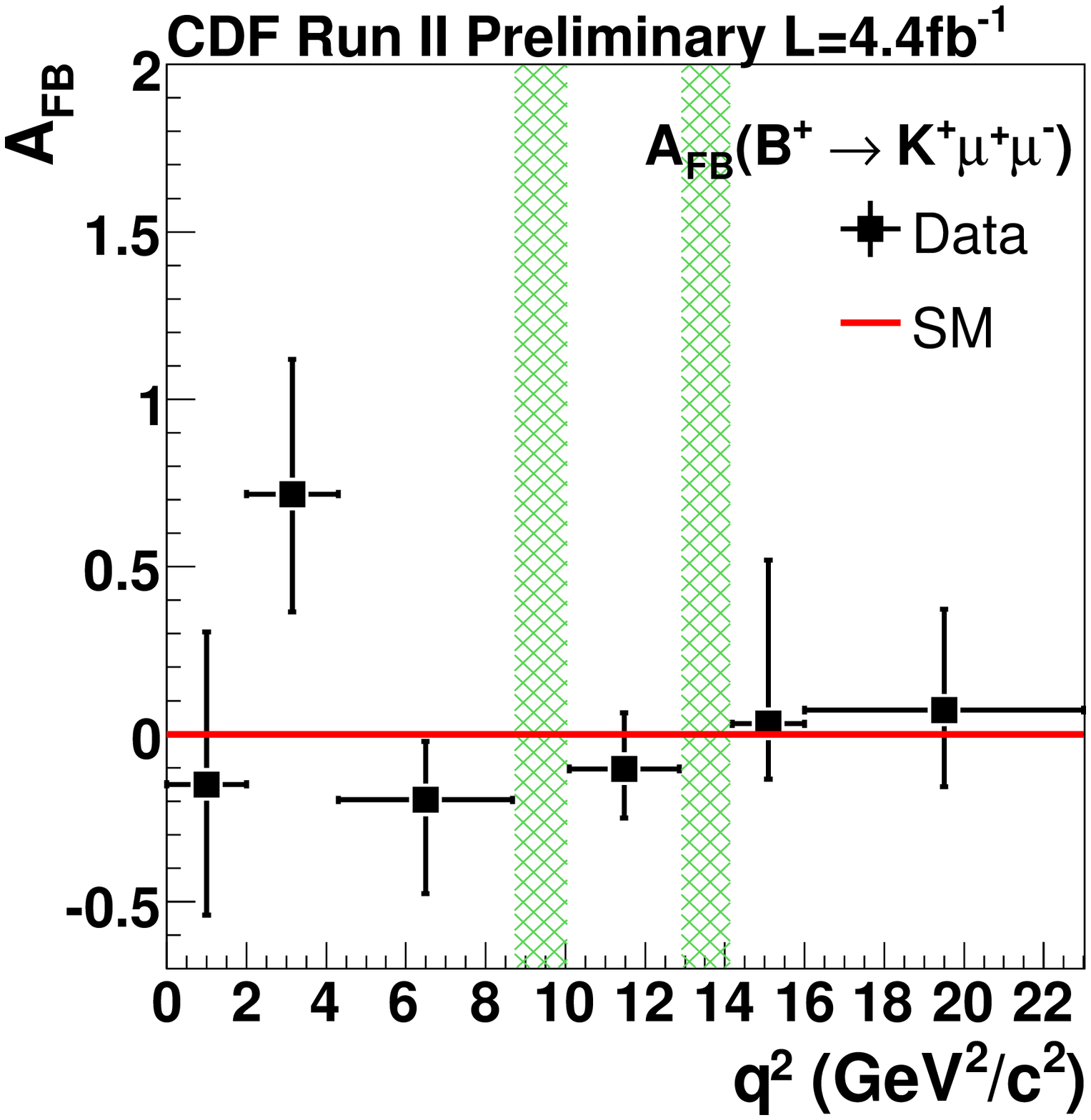}
    \caption{\label{fig:fl_afb} CDF data.
      $F_L$ (left) and $A_{FB}$ (middle) fit results as a function of $q^2$ for $B^0 \to K^{*0} \mu^+ \mu^-$, and $A_{FB}$ (right) as a function of $q^2$ for $B^+ \to K^+ \mu^+ \mu^-$.
      The points show data. Solid (dotted) curve is the SM (an example of SUSY) expectation~\cite{Ali:1999mm}.
      The dashed line is the averaged expectation in each $q^2$ bin.
      The hatched regions are vetoed to prevent contributions from charmonium.
    }
  \end{center}
\end{figure}

CDF reports new results on measurements of $B^+ \to K^+ \mu^+ \mu^-$, $B^0 \to K^{*0} \mu^+ \mu^-$ and $B^0_s \to \phi \mu^+ \mu^-$ decays~\cite{cdf_kmumu}.
Event selection starts from selecting two oppositely charged muon candidates with a transverse momentum greater than $1.5~{\rm GeV}$ or $2.0~{\rm GeV}$ depending on the trigger selection.
$B \to h \mu^+ \mu^-$ candidates are then reconstructed by adding $h = K^+$, $K^{*0}$, or $\phi$ candidates to the dimuon pair.
The $K^{*0}$ is reconstructed in the mode $K^{*0} \to K^+ \pi^-$ and the $\phi$ is reconstructed as $\phi \to K^+ K^-$.
The kaon and pion candidates are required to be consistent with the time-of-flight (TOF) and $dE/dx$ combined log-likelihood probability of each particle hypothesis.
The muon candidates are further purified by the muon-likelihood.
To enhance separation of signal from background, an Artificial Neural Network  multivariate classifier is constructed.
Figure~\ref{fig:bmass_smm} shows the invariant mass distribution for each rare decay.
By performing an unbinned maximum log-likelihood fit of the $B$ invariant mass distribution,
$120 \pm 16$ events for $B^+ \to K^+ \mu^+ \mu^-$, $101 \pm 12$ events for $B^0 \to K^{*0} \mu^+ \mu^-$, and $27 \pm 6$ events for $B^0_s \to \phi \mu^+ \mu^-$ are found,
with $8.5\sigma$, $9.7\sigma$ and $6.3\sigma$ statistical significance, respectively.
This is the first observation of $B^0_s \to \phi \mu^+ \mu^-$ decay.
Using the corresponding $B \to J/\psi h$ modes as a reference, CDF determines the following absolute branching fractions:
${\cal B}(B^+ \to K^+ \mu^+ \mu^-) = [0.38 \pm 0.05 \pm 0.03] \times 10^{-6}$,
${\cal B}(B^0 \to K^{*0} \mu^+ \mu^-) = [1.06 \pm 0.14 \pm 0.09] \times 10^{-6}$ and
${\cal B}(B^0_s \to \phi \mu^+ \mu^-) = [1.44 \pm 0.33 \pm 0.46] \times 10^{-6}$,
where the first uncertainty is statistical and the second is systematic.
The differential branching fractions with respect to $q^2 = m_{\mu\mu}^2 c^2$ for
$B^0 \to K^{*0} \mu^+ \mu^-$ and $B^+ \to K^+ \mu^+ \mu^-$ are also calculated as shown in Fig.~\ref{fig:dbr}.
The forward-backward asymmetry ($A_{FB}$) and $K^{*0}$ longitudinal polarization ($F_L$) are
extracted from $\cos \theta_\mu$ and $\cos \theta_K$ distributions, respectively,
where $\theta_\mu$ is the helicity angle between $\mu^+$ ($\mu^-$) direction
and the direction opposite to the $B$ ($\bar{B}$) direction in the dimuon rest-frame,
and $\theta_K$ is the angle between the kaon direction and the direction opposite to the $B$ meson in the $K^{*0}$ rest frame.
The fit results of $F_L$ and $A_{FB}$  for $B^0 \to K^{*0} \mu^+ \mu^-$ are shown in Fig.~\ref{fig:fl_afb}, as well as
$A_{FB}$ for $B^+ \to K^+ \mu^+ \mu^-$.
Both $F_L$  and $A_{FB}$ are consistent with the SM expectation, and also with an example of a SUSY model~\cite{Ali:1999mm}.
These results are also consistent and competitive with the $B$-factories' measurements~\cite{:2009zv,:2008ju}.


\newpage
\section{Conclusion}

CDF reports the first observation of the decay $B^0_s \to \phi \mu^+ \mu^-$,
results on the forward-backward asymmetry measurements in $B^0 \to K^{*0} \mu^+ \mu^-$ and $B^+ \to K^+ \mu^+ \mu^-$ decays using $4.4~{\rm fb}^{-1}$ of integrated luminosity,
and upper limits on the branching fraction \BRBSMUMU$<4.3\times10^{-8}(3.6 \times 10^{-8})$, ${\cal B}(B^0 \to \mu^+ \mu^-) < 7.6 \times 10^{-9} (6.0 \times 10^{-9})$
at the 95\%~(90\%)~C.L. using $3.7~{\rm fb}^{-1}$ of integrated luminosity.
D0 also reports a new upper limit \BRBSMUMU$<5.1\times10^{-8}(4.2 \times 10^{-8})$ at the 95\%~(90\%)~C.L. using $6.1~{\rm fb}^{-1}$ of integrated luminosity.
There is not a significant discrepancy from the SM expectation in the FCNC $B$ decays so far.
All these results are done with about half or less the statistics expected in a year from now and the analyses are continuously improved,
therefore we could expect large improvements in the near future.

\end{document}